\documentclass[twocolumn,showpacs,preprintnumbers,amsmath,amssymb]{revtex4}
\usepackage{epsfig}
\preprint{NaCoO/2004}
\begin{document}

\title{The infrared spectrum of Na$_{0.57}$CoO$_2$}
\author{S. Lupi, M. Ortolani, and P. Calvani}
\affiliation{Coherentia-INFM and Dipartimento di Fisica, Universit\`a di Roma La Sapienza, Piazzale Aldo Moro 2, I-00185 Roma, Italy}
\date{\today}

\begin{abstract}
The optical conductivity of single crystals of Na$_{x}$CoO$_2$ (for $x$ = 0.57) is first reported between 295 and 30 K. In the far infrared, an "anomalous Drude" analysis leads to a carrier effective mass of 5$\pm 1$ electron masses. That this high value is due to strong electron-phonon coupling is suggested by the Fano distortion of a phonon at 570 cm$^{-1}$. A peak at 8800 cm$^{-1}$ scales with the charge transfer band of several high-$T_c$ cuprates by simply replacing the in-plane Cu-O bond length with that for Co-O.
\end{abstract} 
\maketitle

The layered oxides Na$_{x}$CoO$_2$ have attracted the attention of the scientific community after the discovery that samples with $x \simeq$ 0.3 become superconducting, below $T_c \simeq$ 5 K, after exposure to humid air.\cite{Takada} These hole-doped metals have a hexagonal structure, with the $ab$ planes formed by CoO$_2$ sheets intercalated by layers of Na$^+$ ions. Water molecules between those planes increase the lattice parameter along the $c$-axis from 1.09 to 1.96 nm.\cite{Takada} However, how water inclusion triggers superconductivity in Na$_{x}$CoO$_2$ is not fully understood yet. Therefore, also the investigation of non-hydrated samples is deserving a considerable research effort. Na$_{x}$CoO$_2$ is considered to be a strongly correlated system, with a Hubbard splitting $U \sim$ 5 eV. The conduction band should be narrow, with an effective mass of the carriers $m^* \sim 7m$,\cite{Yang} where $m$ is the bare electron mass. In the $ab$ planes, the resistivity $\rho(T)$ is quasi-linear with $T$ for $x$ = 0.7,\cite{Hasan} with a change of slope around 100 K, while for $x$ = 0.75 $\rho(T)$ is suggestive of a spin ordering instability below 22 K.\cite{Motohashi} For $x$ = 0.57, $\rho(T)$ follows a $T^{3/2}$ law in the $ab$ plane\cite{Rivadulla1} below 175 K. Along the $c$ axis, it has a maximum at the same temperature.\cite{Valla} The latter behavior may be related with the growth, at low temperature, of a coherent quasiparticle peak in the density of states measured by Angle Resolved Photoemission Spectroscopy (ARPES).\cite{Valla} 

In the optical domain, only room-temperature Raman data are presently available to our knowledge: those of Iliev \textit{et al.}\cite{Iliev} on Na$_{0.7}$CoO$_2$ and those of Lemmens \textit{et al.} on Na$_{x}$CoO$_2$$\cdot y$H$_2$O.\cite{Lemmens} Group theory predicts 4$E_{1g}$+4$A_{1g}$ Raman active modes. However, the spectra of dry powders show one $E_{1g}$ phonon at 480 cm$^{-1}$ and one $A_{1g}$ mode at 598 cm$^{-1}$. In the $ab$ planes of single crystals with $x \simeq$ 0.7, one observes a mode at 588 cm$^{-1}$ with weak satellites and, in the hydrated compound, an additional peak at 470 cm$^{-1}$. All these values may change considerably after sample ageing and/or surface treatment.
To our knowledge, no infrared data are presently available for the system Na$_{x}$CoO$_2$. However, they might provide valuable information on the charge dynamics, complementary to that collected by ARPES and transport measurements. The present paper is aimed at filling this gap.

\begin{figure}
{\hbox{\psfig{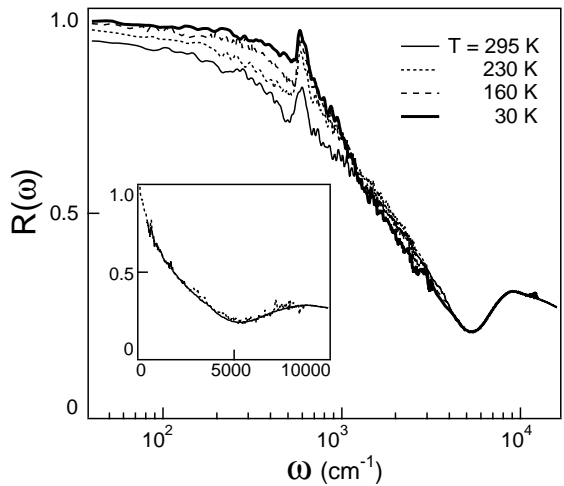}}}
\caption{The reflectivity of a mosaic of coplanar Na$_{0.57}$CoO$_{2}$ crystals at different temperatures, measured with the electric field in the $ab$ plane. In the inset, the $R(\omega)$ of the mosaic at 295 K (dotted line) is compared with that measured on a single crystal by an infrared microscope (solid line).}
\label{R}
\end{figure}

Our experiment was made possible by the use of mosaics of Na$_{0.57}$CoO$_{2}$ single crystals. They were protected from humidity by keeping them in dry air at +45$^0$C, and in a high vacuum when mounted in the cryostat. All of them were grown from the flux as described in Ref. \onlinecite{Rivadulla1} and were extracted from the same batch. After growth, $x$ resulted to be 0.57, the closest value to 0.5 that could be obtained.\cite{Rivadulla1}  Ten to twelve small crystals were mounted on the sample holder. The resulting mosaic surface was approximately 2x2 mm wide. The crystal surfaces, that are excellent $ab$ planes as they are flux-grown, were coplanar within a few degrees. This was obtained by monitoring the reflections of a laser beam focused on the individual crystals and adjusting their positions while the glue (silver paint) was drying. The eventual contribution of the $c$ axis due to residual disalignment of a crystal was largely reduced by the effect of refraction, which bends the radiation wavevector  toward the normal to the surface. As final check, the reflectivity of one crystal was measured at room temperature by an infrared microscope Hyperion 2000, mounted onto a Bruker IFS 66V interferometer, between 400 and 17,000 cm$^{-1}$. The reference was the crystal itself after exposure to gold evaporation. The resulting $R(\omega)$ (solid line) is compared with that of the mosaic at the same temperature (dotted line) in the inset of Fig.\ \ref{R}.

The $R(\omega)$ of the mosaic was measured between 30 and 295 K, and between 40 and 10000 cm$^{-1}$ at quasi-normal incidence (8$^0$), by a Bomem DA3 interferometer. The final alignment of the reflectivity setup was performed by remotely controlled motors, in order to compensate for the mechanical stress of the interferometer optics after the evacuation. The reference was a gold film evaporated {\it in situ} onto the sample by displacing a hot filament in front of it. The $R(\omega)$ thus measured is shown at different temperatures in Fig.\ \ref{R}. 
The complex optical conductivity $\tilde \sigma(\omega)$ was then obtained by standard Kramers-Kronig relations. A Hagen-Rubens extrapolation was employed below 40 cm$^{-1}$, an extrapolation based on fits by Lorentzians and polynomials was used above 17,000 cm$^{-1}$. An alternative extrapolation based on the reflectivity\cite{Terasaki} of NdCoO$_{3}$ gave the same results within errors. As the surface of the flux-grown crystals is intrinsically $ab$, the $\tilde \sigma(\omega)$ obtained in this way is the optical conductivity of the CoO$_2$ sheets of Na$_{0.57}$CoO$_{2}$.

\begin{figure}
{\hbox{\psfig{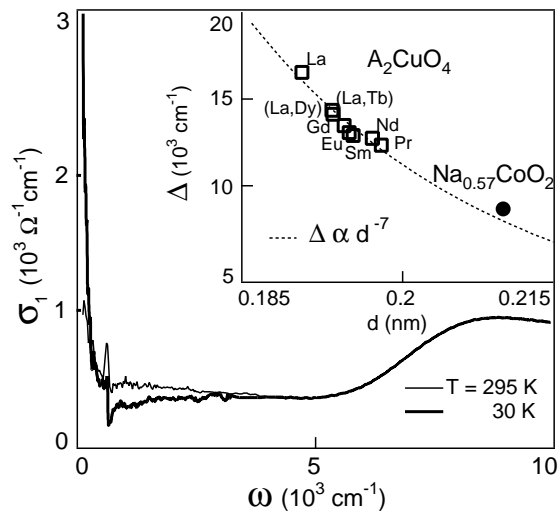}}}
\caption{$\sigma(\omega)$ at different temperatures, as extracted from the reflectivity of Fig. 1. In the inset, the energy $\Delta$ of the peak here observed in the near infrared (full circle) is compared with those reported in Ref. 10 for the charge-transfer bands of several cuprates belonging to the family A$_{2}$CuO$_{4}$, with A = La, (La,Tb), (La,Dy), Gd, Nd, Eu, Sm, Pr (squares). The dashed line is a $\Delta \propto d^{-7}$ law, where $d$ is the $ab$ plane Co-O, or Cu-O distance, respectively.}
\label{sigma}
\end{figure}

The real part of $\tilde \sigma(\omega)$, $\sigma_1(\omega)$, is shown at two temperatures in Fig.\ \ref{sigma}. In both sets of data, the main features are a metallic conductivity in the far and mid infrared, a phonon line, and a broad absorption in the near infrared. This latter exhibits an edge around 6000 cm$^{-1}$ and a well defined peak at $\Delta$ = 8800 cm$^{-1}$. This energy roughly corresponds to that (0.7 eV) of a peak observed in ARPES spectra, and assigned to Co $t_{2g}$-derived states.\cite{Hasan,Yang} However, the infrared band at 8800 cm$^{-1}$ can be rather attributed to charge transfer (CT) between the Co $3d$ orbitals and the O $2p$ orbitals. This assignment is supported by scaling the energy of the band of Na$_{0.57}$CoO$_{2}$ with that of the corresponding bands of the cuprates. In high-$T_c$ materials, where the Cu-O bonds of length $d$ are hybridized $p-d$ and have a octahedral coordination, the CT peak energy $\Delta$ is found to follow a law $\Delta \propto d^{-7}$.\cite{Tokura,Cooper} The same hybridization and coordination are present in Na$_{x}$CoO$_{2}$, despite its unit cell being hexagonal instead of tetragonal. As shown in the inset of Fig.\ \ref{sigma}, the present value of $\Delta$ for the CoO$_2$ sheets (where $d_{Co-O}$  = 0.210 nm) is in excellent agreement with the prediction obtained by extrapolating the $d^{-7}$ law which holds for cuprates. The latter compounds in Fig.\ \ref{sigma} are undoped, unlike Na$_{0.57}$CoO$_{2}$. However, it has been shown that the energy of the charge-transfer band in La$_{2-x}$Sr$_x$CuO$_{4}$ is approximately independent of doping.\cite{Lucarelli}

\begin{figure}
{\hbox{\psfig{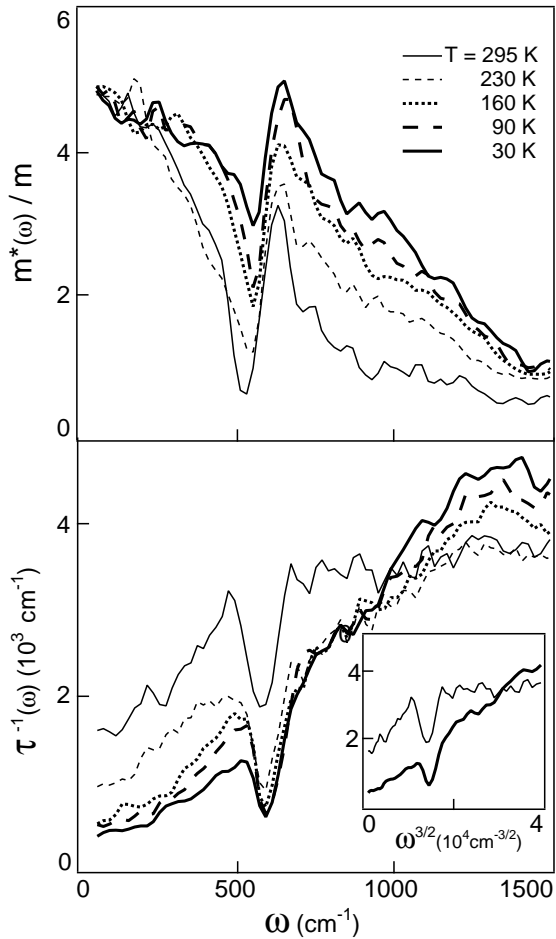}}}
\caption{Effective mass (top) and relaxation rate (bottom) vs. frequency, as obtained from $\tilde \sigma(\omega)$ of Na$_{0.57}$CoO$_{2}$. In the inset, $\tau^{-1}(\omega)$ is plotted vs. $\omega^{3/2}$ both at room temperature (thin line) and at 30 K (thick line).}
\label{anomalous}
\end{figure}

The conductivity of oxides within the charge-transfer gap has been often described in terms of an "anomalous-Drude" model, where both the effective mass $m^*(\omega)$ and the relaxation rate $\tau^{-1}(\omega)$ of the carriers are allowed to depend on frequency. Both those quantities, as well as the plasma frequency $\omega_{p}$, can be directly extracted\cite{Puchkov}  from $\tilde \sigma(\omega)$. The results for Na$_{0.57}$CoO$_{2}$ are shown in Fig. \ \ref{anomalous}. The strong phonon absorption around 570 cm$^{-1}$, that will be discussed later on, does not prevent one from obtaining information on the carrier parameters.  First, for $\omega \alt$ 1500 cm$^{-1}$, an "anomalous Drude" regime indeed holds, where both $m^*/m$ and $\tau^{-1}$ depend on $\omega$. There is no indication for the opening of a pseudogap\cite{Puchkov} down to 30 K, which would suggest either charge localization or pair pre-formation. Secondly, at any $T$, $m^*(\omega)/m \to 5 \pm 1$ for $\omega \to 0$, in satisfactory agreement with the photoemission data cited above.\cite{Yang} Thirdly, below a temperature $T_0$ between 230 and 160 K, $\tau^{-1} \propto \omega^{3/2}$ (see the inset of Fig. \ \ref{anomalous}). Moreover, below $T_0$ its low-frequency values become independent of $T$ within errors. This is clearly suggestive of a change in the transport properties and may be related to the crossover observed at $T_M \sim$ 180 K both in the ARPES spectra\cite{Valla} of Na$_{0.57}$CoO$_{2}$ and in its resistivity,\cite{Rivadulla1,Valla} which is proportional to $T^{3/2}$ below $T_M$.

The $E_{1u}$ optical phonon observed around 570 cm$^{-1}$ deserves a detailed description. Four of such modes are predicted by group analysis but, as for Raman spectra cited above,\cite{Iliev} just one line is observed here. It can be assigned to the Co-O stretching,\cite{Sudheendra} consistently with its strong hardening for decreasing temperature (see $\widetilde\omega_F$ in Table I). In the present infrared spectra, the other phonon lines, which are expected to occur at lower frequencies than the Co-O stretching, are probably shielded by the narrow free-carrier continuum in a more efficient way. 

\begin{figure}
{\hbox{\psfig{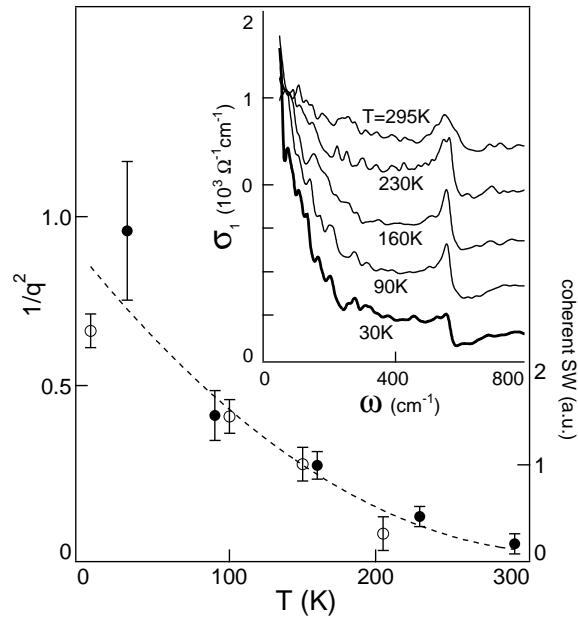}}}
\caption{The Fano parameter $1/q^2$ of the phonon at 570 cm$^{-1}$ (dots) is reported vs. $T$ and compared with the spectral weight SW of the coherent ARPES peak in Ref. 5 (circles). The SW at 150 K has been normalized to the $1/q^2$ value at 160 K. The dashed line is a guide to the eye. In the inset, the observed Fano profile is shown at different temperatures. An offset has been introduced for sake of clarity.}
\label{q-sw}
\end{figure}

The phonon line around 570 cm$^{-1}$, as shown in the inset of Fig.\ \ref{q-sw}, exhibits a pronounced Fano distortion. This indicates a strong coupling between the Co-O stretching and the Drude-like continuum, which indeed has most of its spectral weight on the opposite side of the absorption minimum.\cite{Lupi98} The effect is more and more evident as $T$ is lowered. The observed spectrum can be accurately fitted at all temperatures by the Fano equation for the dielectric function\cite{Lupi98,Davis}

\begin{equation}
\epsilon_{2F }(\omega) = 
                   R\left({{(q\gamma_F+(\omega-\widetilde\omega_F))^2}
                   \over{\gamma_F^2+(\omega-\widetilde\omega_F)^2}}-1\right). 
\label{Fano}
\end{equation}

\noindent 
Therein, $\widetilde\omega_F$ is the renormalized frequency of the phonon and $\gamma$ is its half width, which here basically measures the electron-phonon interaction strength. $R$ is a scale parameter which includes the transition rates, and $q$ is the Fano-Breit-Wigner parameter, which measures 
the reciprocal of the electron-phonon interaction once weighted for the density of states of the continuum.\cite{Fano61} 

The quantities in Eq.\ \ref{Fano}, as provided by the fit to data, are reported in Table I. One may notice that $\gamma_F$, which measures the electron-phonon interaction, is large and independent of temperature within the errors reported in parenthesis. On the other hand, as already mentioned, $\widetilde\omega_F$ appreciably increases as $T \to 0$. Also $1/q^2$ steadily increases for decreasing $T$ as shown in Fig.\ \ref{q-sw}. Therein, its behavior is show to be consistent with that of the spectral weight of the coherent peak which appears in the ARPES spectra, close to the Fermi level, for $T <$ 200 K. Indeed, as the electron-phonon interaction $\gamma_F$ is $T$-independent in Table I, and the same holds for the matrix element of the dipolar Hamiltonian, the $T$-dependence of $1/q^2$ probes that of the density of states of the continuum at the phonon energy. 


\begin{table}
\caption{Values of the Fano parameters in Eq. 1, as provided by the fit to data at five temperatures. An estimate of the average experimental uncertainty, which holds for all temperatures, is given in parenthesis.}
\begin{ruledtabular}
\begin{tabular}{cccccc}
$T$ (K) & 295 & 230 & 160 & 90 & 30 \\
q          & -5.7(0.2) & -3.1 & -1.9 & -1.6 & -1.0 \\
R          & 1.0(0.3)  & 6.9 & 14.6 & 18.2 & 21.3 \\          
$\gamma_F$ (cm$^{-1}$) & 21(5)  & 25 & 18 & 19 & 26 \\
$\widetilde \omega_F$ (cm$^{-1}$) & 562(2)  & 563 & 566 & 569 & 575 \\
\label{TABLE I}
\end{tabular}
\end{ruledtabular}
\end{table}

In conclusion, we have reported here the reflectivity spectra of single crystals of Na$_{0.57}$CoO$_2$ in the $ab$ plane. The resulting optical conductivity exhibits a near infrared peak that can be attributed to Co(3d)-O(2p) charge transfer. Indeed, it scales with that charge transfer band observed in the cuprates by simply replacing the Cu-O in-plane bond length with the Co-O bond length. In the far infrared, $\tilde \sigma(\omega)$ can be described by an anomalous Drude behavior, where the carriers have an effective mass of about 5$m$. Their relaxation rate varies as $\omega^{3/2}$ and is independent of $T$, within errors, below a temperature $T_0$. $T_0$ is consistent with the crossover temperature $T_M$ below which either $\rho(T) \propto  T^{3/2}$ and a quasiparticle peak appears in the photoemission spectra of Na$_{0.57}$CoO$_2$. 

Only one unshielded phonon has been observed in the present spectra, namely the transverse optical Co-O stretching mode. It exhibits a marked Fano profile, which is suggestive of a strong electron-phonon coupling. This is consistent with the high effective mass of the carriers, which therefore could be polaronic in nature. Under this assumption, the observation of the quasiparticle photoemission peak below $T_M$  and, here, of a $\tau^{-1}$ independent of $T$ below $T_0$, may be interpreted in terms of coherent (tunneling) propagation of large polarons. On the other hand, even if we observe a broad background in the midinfrared, which cannot be described in terms of Drude-like models, we cannot identify any clear polaronic "shake-off" band in the present data. Systematic infrared investigations of the Na$_{x}$CoO$_2$ system as a function of doping and temperature would be needed to clearly identify the nature of the carriers.

We wish to thank F. Rivadulla for producing and characterizing the samples here measured, as well as for many valuable discussions.

%
%

\noindent

\end{document}